\documentclass[prl,reprint,amsmath,amssymb,amsfonts,showpacs,floats]{revtex4-1}

\usepackage{graphicx}
\usepackage{multirow}
\usepackage{color}
\usepackage{amsthm}
\usepackage{ulem}
\usepackage{enumerate}
\usepackage{rotating}
\usepackage[table]{xcolor}
\usepackage{hyperref} 

\graphicspath{{../plots/}}

\newcommand{\mc}{\mathcal}

\newcommand{\nn}{\nonumber}

\newcommand\redout{\bgroup\markoverwith
{\textcolor{red}{\rule[.5ex]{2pt}{.4pt}}}\ULon}

\begin{document}

\title{Nonlocal Entanglement of 1D Thermal States Induced by Fermion Exchange Statistics}

\author{YeJe Park}
\author{Jeongmin Shim}
\author{S.-S. B. Lee} \email[Present address: Physics Department, Arnold Sommerfeld Center for Theoretical Physics, and Center for NanoScience, Ludwig-Maximilians-Universit\"{a}t, Theresienstra{\ss}e 37, D-80333 M\"{u}nchen, Germany]{}
\author{H.-S. Sim}\email[]{hssim@kaist.ac.kr}
\affiliation{Department of Physics, Korea Advanced Institute of Science and Technology, Daejeon 34141, Korea}

\begin{abstract}
When two identical fermions exchange their positions, their wave function gains phase factor $-1$.
We show that this distance-independent effect can induce nonlocal entanglement in one-dimensional (1D) electron systems having Majorana fermions at the ends. It occurs in the system bulk and has nontrivial temperature dependence. In a system having a single Majorana at each end, the nonlocal entanglement has a Bell-state form at zero temperature and decays as temperature increases, vanishing suddenly at certain finite temperature. In a system having two Majoranas at each end, it is in a cluster-state form and its nonlocality is more noticeable at finite temperature. By contrast, thermal states of corresponding 1D spins do not have nonlocal entanglement.  
\end{abstract}

\date{\today}



\maketitle

Topological phases of matter are notions of zero-temperature ground states.
They have interesting properties such as ground-state degeneracy, edge states, and unusual excitations.
For example, topological phases of 1D electron systems have Majorana zero modes
localized at system ends~\cite{kitaev01,Fidkowski11,Turner11,Wen12}. The phases are identified by entanglement spectrum~\cite{Fidkowski11,Turner11}, entanglement entropy~\cite{Kim14}, or a nonlocal string order parameter~\cite{Bahri14}, and classified by symmetries and the number of the zero modes~\cite{Fidkowski11,Turner11,Wen12}. The non-Abelian statistics of Majorana fermions is a key to topological quantum computing~\cite{Kitaev03,Alicea11}.

A natural question is how quantum properties of topological phases are thermally suppressed~\cite{Castelnovo07,Nussinov08,Hastings11,Poulin13,Viyuela14,Arovas14,Huber14,Budich15,Grusdt16}.
It involves a number of largely unexplored points. 
First of all, tools identifying topological phases of pure states are not readily applicable to thermal states;  e.g. for mixed states entanglement spectrum~\cite{Li08,Fidkowski10,Pollmann10} is ill-defined and entanglement entropy overestimates entanglement.
Secondly, pure ground states have definite quantum numbers in presence of symmetries, such as number parity of 1D electrons, and they can be thermally mixed. Consequences of such mixing need to be studied.
Lastly, topological phases of 1D electron pure states can be classified by means of the 1D spins obtained by Jordan-Wigner transformation (JWT)~\cite{Chen}.
It is interesting to see whether this fermion-spin correspondence is applicable to thermal states.
We will address the above points for 1D electrons.


In this Letter, we study thermal states of 1D electron systems having Majorana zero modes, using two mixed-state entanglement measures~\cite{Plenio07}, the entanglement of formation~\cite{Bennett96} and logarithmic negativity~\cite{Lee00,Vidal02,Plenio05}.
We find that the thermal states 
can have 
{\it nonlocal} and {\it length-independent} entanglement of occupation number in the {\it bulk}.
It has nontrivial temperature ($T$) dependence, depending on the number of the zero modes. In a system having one zero mode at each end, nonlocal entanglement occurs in a Bell-state form at $T=0$, decreases at larger $T$, then vanishes at certain finite $T$.
In a system of two zero modes at each end, it occurs in a cluster-state form and its nonlocality is noticeable in a finite-$T$ window.
The nonlocal entanglement results from the non-Abelian statistics of Majorana fermions and the fermion exchange statistics.
By contrast, the 1D spins, obtained by JWT of the 1D electrons, do not have any nonlocal entanglement. 
 
\begin{figure}[b]
\includegraphics[width=\columnwidth]{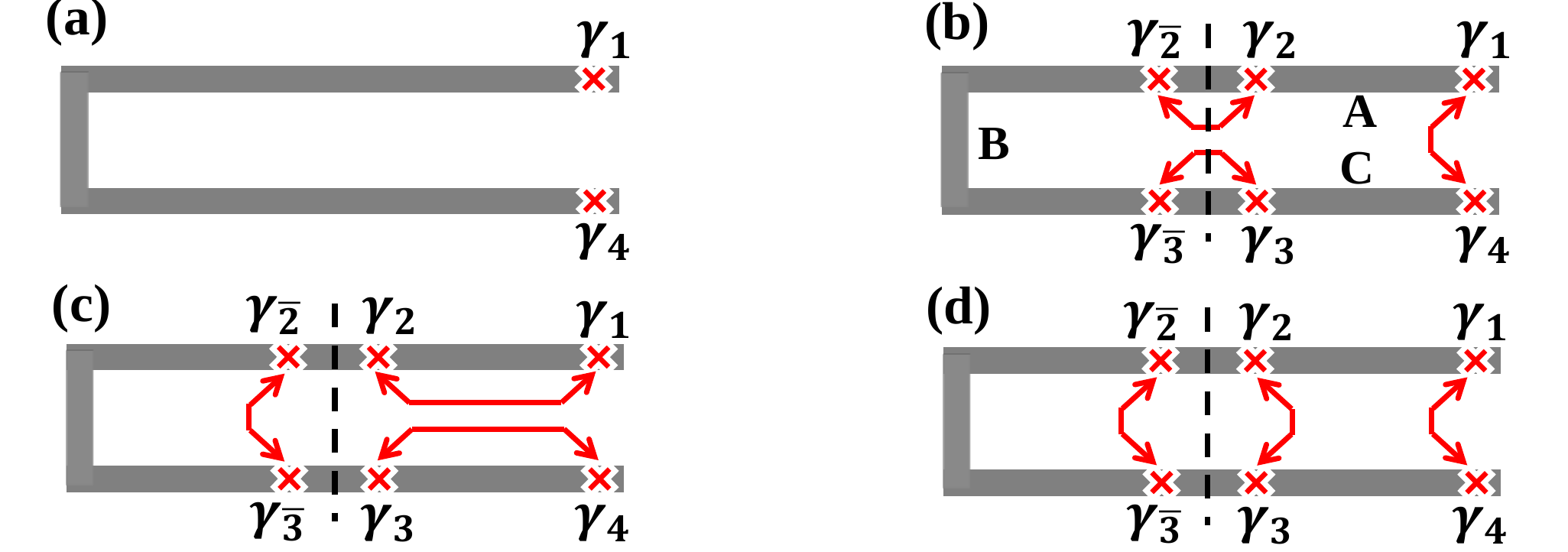}
\caption{ (a) Kitaev chain having one Majorana zero mode $\gamma_{1,4}$ at each end. (b) It is partitioned into the left (B) and right (A and C) by dashed imaginary cuts. There are Majorana fermions $\gamma_{2,\bar{2},\bar{3},3}$ at the cuts. Majorana fusion pairs (arrows) $\gamma_2 + i \gamma_{\bar{2}}$ and $\gamma_{\bar{3}} + i \gamma_3$ annihilate the ground states of the chain. A fusion pair $\gamma_{\bar{2}} + i \gamma_{\bar{3}}$ is localized within the left, while fusion pairs (c) $\gamma_1 + i \gamma_2$, $\gamma_3 + i \gamma_4$ or (d) $\gamma_2 + i \gamma_3$, $\gamma_1 + i \gamma_4$ in the right. } \label{system_H1}
\end{figure}
 
{\it Nonlocal entanglement in Kitaev chain.---} We first consider a Kitaev chain~\cite{kitaev01} of spinless electrons. 
It represents a topological phase (BDI class) of 1D fermions protected by time-reversal and parity symmetries~\cite{Fidkowski11, Turner11, Wen12}. 
It has one Majorana zero mode $\gamma_{a=1,4}$ at each end [Fig.~\ref{system_H1}(a)]. We study entanglement between the left B and right AC of imaginary cuts through physical bonds connecting electron sites [Fig.~\ref{system_H1}(b)], partitioning the chain into regions A, B, C. 
We use fermion operators $f_{ab} \equiv (\gamma_a + i \gamma_{b}) / \sqrt{2}$ and their occupation numbers $\hat{n}_{ab} = f^\dagger_{ab} f_{ab}$, formed by Majoranas at the ends $\gamma_{a=1,4}$ and the cuts  $\gamma_{a=2,\bar{2},\bar{3},3}$.
The system Hamiltonian is  
$\hat H_\textrm{I} = - \sum_{j=1}^{N-1} [\frac{t}{2} (c_j^{\dagger} c_{j+1} + c_{j+1}^{\dagger} c_j) + \frac{\Delta}{2} (c_j c_{j+1} + c^\dagger_{j+1} c^\dagger_j )] +\sum_{j=1}^N \mu c_j^{\dagger} c_j$ 
with hopping energy $t$,  
pairing $\Delta$, chemical potential $\mu$, and electron creation $c_j^\dagger$ at site $j =1,\cdots,N$.
We choose~\cite{Other} $t = \Delta>0$ and $\mu = 0$ for simplicity.
Then the chain has two degenerate ground states, $|0 \rangle_\textrm{I} = |0_{14} 0_{2\bar{2}} 0_{\bar{3}3} \cdots \rangle$ (of even number parity) and $|1 \rangle_\textrm{I} = f^\dagger_{14} |0 \rangle_\textrm{I}$ (odd).
At equilibrium, it is described by the thermal state 
$\rho_\textrm{I} (T) = e^{- \beta \hat{H}_\textrm{I}}/\textrm{Tr} e^{- \beta \hat{H}_\textrm{I}}$ with $\beta = k_B T$ and Boltzmann constant $k_B$.
 
We analyze entanglement between B and AC in $\rho_\textrm{I} (T=0) = (|0 \rangle \langle 0 |_\textrm{I} + |1 \rangle \langle 1 |_\textrm{I})/2$ at $T=0$. For this, we write~\cite{Supple} the ground states using operators $f_{12}$, $f_{\bar{2}\bar{3}}$, $f_{34}$ localized within A, B, C, respectively [Fig.~\ref{system_H1}(c)],
\begin{align} 
|0\rangle_\textrm{I} = \frac{1}{2}  ( & 1 + f_{12}^\dagger f_{34}^\dagger + f_{12}^\dagger f_{\bar{2}\bar{3}}^\dagger    + f_{\bar{2} \bar{3}}^\dagger f_{34}^\dagger  )|0_{12}   0_{\bar{2}\bar{3}}   0_{34} \cdot \cdot \rangle,  \nonumber \\
|1\rangle_\textrm{I} = \frac{1}{2}( &
 f_{12}^\dagger + f_{34}^\dagger + f_{\bar{2}\bar{3}}^\dagger  +  f_{12}^\dagger f_{\bar{2}\bar{3}}^\dagger f_{34}^\dagger ) |0_{12}   0_{\bar{2}\bar{3}}   0_{34} \cdot \cdot \rangle. \label{ABCstates} 
\end{align}
To see the entanglement, we need to map~\cite{Zanardi02,Banuls07} them into qubits since fermion states lack tensor product structure.
Before mapping, we reorder operators such that those belonging to a subsystem are grouped together;
$|0 \rangle_\textrm{I} = \frac{1}{2} (1 + f_{12}^\dagger f_{34}^\dagger - f_{\bar{2}\bar{3}}^\dagger f_{12}^\dagger + f_{\bar{2} \bar{3}}^\dagger f_{34}^\dagger  )|0_{\bar{2}\bar{3}} 0_{12} 0_{34} \cdots \rangle$, $|1 \rangle_\textrm{I} = \frac{1}{2} (f_{12}^\dagger + f_{34}^\dagger + f_{\bar{2}\bar{3}}^\dagger - f_{\bar{2}\bar{3}}^\dagger f_{12}^\dagger f_{34}^\dagger ) |0_{\bar{2}\bar{3}} 0_{12} 0_{34} \cdots \rangle$.
Here, $f_{12}^\dagger$ and $f_{34}^\dagger$ belonging to AC are collected to the right of $f_{\bar{2}\bar{3}}^\dagger$ belonging to B, generating the fermion exchange sign $-1$ in some coefficients.
We call this ordering as ``subsystem operator grouping''. Then occupation number states are mapped onto qubit states,
\begin{align}
| 0\rangle_\textrm{I} \mapsto |0\rangle_\textrm I^q= & \frac{1}{2}(| 0_{\bar{2}\bar{3}} \rangle^q (|0_{12}\rangle^q |0_{34} \rangle^q +|1_{12} \rangle^q  |1_{34} \rangle^q) \nonumber \\
& - |1_{\bar{2}\bar{3}} \rangle^q (|1_{12} \rangle^q  |0_{34} \rangle^q -|0_{12}\rangle^q | 1_{34} \rangle^q)), \nonumber \\
|1\rangle_\textrm{I} \mapsto |1\rangle_\textrm I^q= & \frac{1}{2} ( | 0_{\bar{2}\bar{3}}\rangle^q (|1_{12} \rangle^q | 0_{34} \rangle^q +|0_{12} \rangle^q   |1_{34} \rangle^q) \nonumber \\
& + |1_{\bar{2}\bar{3}}\rangle^q (|0_{12} \rangle^q | 0_{34} \rangle^q -|1_{12} \rangle^q |1_{34} \rangle^q)). \label{H1entanglement2}
\end{align}
This shows that $\rho_\textrm{I} (T=0)$ has maximal entanglement between B ($\bar{2} \bar{3}$) and AC (12,34).

This entanglement can be alternatively seen writing the states using $\{f_{23}, f_{14}\}$ instead of $\{f_{12}, f_{34}\}$. 
The result~\cite{Supple} is
$|0 \rangle_\textrm{I} = \frac{1}{\sqrt{2}}(1 + i f^\dagger_{\bar{2} \bar{3}} f^\dagger_{23}) |0_{\bar{2} \bar{3}} 0_{23} 0_{14} \cdots \rangle$, $|1 \rangle_\textrm{I} = \frac{1}{\sqrt{2}} (1 + i f^\dagger_{\bar{2} \bar{3}} f^\dagger_{23}) f^\dagger_{14} |0_{\bar{2} \bar{3}} 0_{23} 0_{14} \cdots \rangle$. We map these onto qubits,
$|0\rangle_\textrm I \mapsto |\text{Bell}\rangle^q |0_{14}\rangle^q$ and $|1\rangle_\textrm I \mapsto |\text{Bell}\rangle^q |1_{14}\rangle^q$. The map shows a Bell state entangling B ($\bar{2}\bar{3}$) and AC (23)
\begin{align}
  |\text{Bell}\rangle^q &= \frac{1}{\sqrt{2}} (|0_{\bar{2}\bar{3}}\rangle^q |0_{23}\rangle^q + i |1_{\bar{2}\bar{3}} \rangle^q |1_{23}\rangle^q). \label{H1entanglement}
\end{align}
Their mixture also has the Bell state, $\rho_\textrm{I}(T=0) \mapsto \rho_\textrm{I}^q(T=0) = |\text{Bell}\rangle \langle \text{Bell}|^q \otimes (|0_{14} \rangle \langle 0_{14}|^q + |1_{14} \rangle \langle 1_{14}|^q)/2$. 
The end qubit $|n_{14} \rangle^q$ does not affect the entanglement.

The Bell state $|\text{Bell} \rangle^q$ is nonlocal as $f^\dagger_{\bar{2} \bar{3}}$ and $f^\dagger_{23}$ create nonlocal fermions.
It originates~\cite{Ivanov01} from entanglement generation
$|0_{2\bar{2}} 0_{\bar{3}3} \rangle \to |0_{\bar{2}\bar{3}} 0_{23} \rangle + i |1_{\bar{2} \bar{3}} 1_{23} \rangle$ by changing 
non-Abelian Majorana fusion pairs from $\gamma_2 + i\gamma_{\bar{2}}$, $\gamma_{\bar{3}} + i \gamma_3$ [Fig.~\ref{system_H1}(b)] to $\gamma_{\bar{2}} + i \gamma_{\bar{3}}$, $\gamma_2 + i \gamma_3$ [Fig.~\ref{system_H1}(d)]. 
It occurs in the chain bulk and is independent of lengths of A, B, C. 
It is unaffected by parity mixing as it occurs in both $|0\rangle_\textrm{I}$ and $|1\rangle_\textrm{I}$. 
It is robust against quasiparticle poisoning~\cite{Goldstein11,Budich12,Rainis12} due to protection by the gap $\Delta$. This can be seen as a {\it bulk-edge correspondence} of the Kitaev chain; the entanglement occurs in the bulk for other values of  $\mu$, $t$, and $\Delta$~\cite{Other} for which end Majoranas appear.

We emphasize the importance of the fermion exchange sign in the mapping~\eqref{H1entanglement2}. As the entanglement is physical, it should be invariant under basis change between $\{ f_{12}, f_{34} \}$ and $\{f_{14}, f_{23} \}$. However, if we were to map $|n=0,1\rangle_\textrm{I}$ into qubits without the subsystem operator grouping, we would obtain the states of Eq.~\eqref{H1entanglement2} but with positive signs replacing the negative signs, 
\begin{align}
| 0\rangle^s_\textrm{I} = & \frac{1}{2}(| 0_{\bar{2}\bar{3}} \rangle^s (|0_{12}\rangle^s |0_{34} \rangle^s +|1_{12} \rangle^s  |1_{34} \rangle^s) \nonumber \\
& + |1_{\bar{2}\bar{3}} \rangle^s (|1_{12} \rangle^s  |0_{34} \rangle^s +|0_{12}\rangle^s | 1_{34} \rangle^s)), \nonumber \\
|1\rangle^s_\textrm{I} = & \frac{1}{2} ( | 0_{\bar{2}\bar{3}}\rangle^s (|1_{12} \rangle^s | 0_{34} \rangle^s +|0_{12} \rangle^s   |1_{34} \rangle^s) \nonumber \\
& + |1_{\bar{2}\bar{3}}\rangle^s (|0_{12} \rangle^s | 0_{34} \rangle^s +|1_{12} \rangle^s |1_{34} \rangle^s)), \label{ABCstates_spin}
\end{align}
where superscripts $s$ are for the discussion about spins below.
Although $|n \rangle^s_\textrm{I}$'s have entanglement between B and AC similar to $|n \rangle_\textrm{I}^q$ in Eq.~\eqref{H1entanglement2}, their equal mixture $\rho_\textrm{I}^s(T=0)=(|0 \rangle \langle 0|^s + |1 \rangle \langle 1 |^s)/2$ has no entanglement in contradiction with the presence of $|\textrm{Bell}\rangle^q$ in $\rho_\textrm{I}(T=0)$ in Eq.~\eqref{H1entanglement}. 
This demonstrates that the fermion statistics has to be taken into account in order to correctly quantify entanglement in fermion mixed states.

We notice that $\rho_\textrm{I} (T=0)$ provides a good example resolving an issue~\cite{Montero11,Friis13} of fermion entanglement whether particular orderings of fermions have to be chosen when one maps fermion occupations onto qubits.


To see whether the fermion-spin correspondence~\cite{Chen, Fidkowski11} is valid for the mixed state $\rho_\textrm{I} (T=0)$, we compare the Kitaev chain with the spin chain obtained by JWT~\cite{Supple} of $\hat{H}_\textrm{I}$.
The ground states $|n \rangle^s_\textrm{I}$ of the spin chain turn out to be the same as those in Eq.~\eqref{ABCstates_spin} obtained from the electron states $|n \rangle_\textrm{I}$ by mapping into qubits without taking the fermion exchange sign into account. $|n \rangle^s_\textrm{I}$'s are eigenstates of JWT of the parity operator  $\hat{P} = \prod_{j} e^{i \pi c^\dagger_j c_j}$ with eigenvalue $(-1)^n$ so that the parity symmetry is mapped by JWT~\cite{Chen}.
In Eq.~\eqref{ABCstates_spin} $|n_{ab}=0,1 \rangle^s$ 
is a state of region R ($= \textrm{A}$ for $ab=12$, $\textrm{B}$ for $\bar{2}\bar{3}$, $\textrm{C}$ for $34$) of the spin chain~\cite{Supple}. 
The pure ground states in Eq.~\eqref{ABCstates_spin} have the same entanglement as those in Eq.~\eqref{H1entanglement2}, as expected from the correspondence. 
However, $\rho_\textrm{I}^s(T=0)$ has no entanglement, showing that the correspondence breaks down for mixed states.

{\it Nonlocal entanglement at $T>0$.---} We quantify entanglement between B and AC in the thermal state $\rho_\textrm{I} (T)$ by two bipartite mixed-state measures, logarithmic negativity $\mathcal{LN}$~\cite{Lee00,Vidal02,Plenio05} and the entanglement of formation $\mathcal{F}$~\cite{Bennett96}. $\mathcal{LN}$ is easier to compute than other measures. 
We compute $\mathcal{LN} (\rho_\textrm{I}) \equiv \log_2 \textrm{Tr} \left|(\rho_\textrm{I}^q)^{T_\textrm{B}}\right|$. $(\rho_\textrm{I}^q)^{T_\textrm{B}}$ is the partial transpose with respect to B of $\rho^q_\textrm I$ obtained by mapping $\rho_\textrm{I}$ onto qubits with the subsystem operator grouping.  $\textrm{Tr} |\cdot |$ is the trace norm. 
$\mathcal{F}$ is a mixed-state generalization of entanglement entropy $\mathcal{S}$, as $\mathcal{F}= \mathcal{S}$ for pure states.
We compute $ \mathcal{F} (\rho_\textrm{I}) \equiv \inf_{\rho^q_\textrm{I} = \sum_i w_k |\psi_k \rangle \langle \psi_k |} [ \sum_k w_k \, \mathcal{S} (|\psi_k \rangle) ]$, exploring all possible decompositions of $\rho^q_\textrm{I}$ into normalized pure states $|\psi_k \rangle$ with weight $w_k$ and finding the optimal decomposition for which $\sum_k w_k \, \mathcal{S} (|\psi_k \rangle)$ is the lowest. $\mathcal{S} (|\psi_k \rangle) \equiv - \textrm{Tr} ( \rho^\textrm{B}_k \log_2 \rho^\textrm{B}_k )$ is entanglement entropy between B and AC in $|\psi_k \rangle$ and $\rho^\textrm{B}_k = \textrm{Tr}_\textrm{B} |\psi_k \rangle \langle \psi_k|$. $\mathcal{LN}$ and $\mathcal{F}$ have been used for studying many-body states~\cite{Bayat12, YALee13, Castelnovo13, Calabrese15, Lee15, Eisler, Shapourian}. 


\begin{figure}[bt]
\includegraphics[width=\columnwidth]{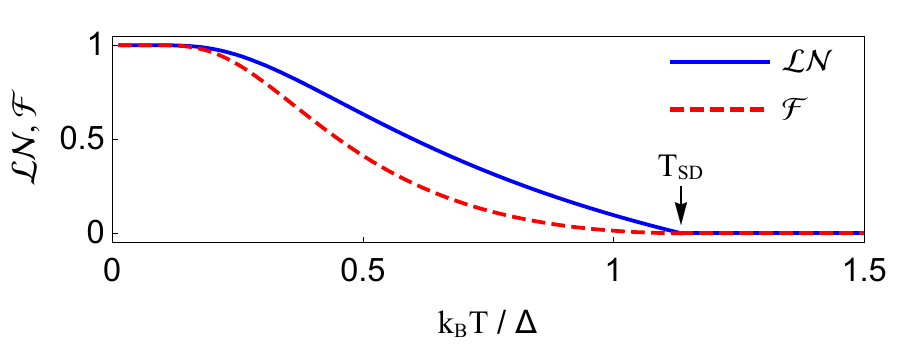}
\caption{Temperature dependence of nonlocal entanglement between the regions B and AC of Kitaev chain. It is quantified by $\mathcal{LN}$ and $\mathcal{F}$. It is maximal at $T=0$ and survives up to $T_\textrm{SD}$. It decays not exponentially nor algebraically near $T_\textrm{SD}$. 
}
\label{H1_finiteT}
\end{figure}

We can write $\rho_\textrm{I} (T) = \rho_\textrm{I,cuts} (T) \rho'_\textrm{I} (T)$. 
The first factor $\rho_\textrm{I,cuts} \propto e^{-\beta \hat H_\textrm{I,cuts}}$ accounts for thermal excitations at the cuts, where $\hat H_\textrm{I,cuts} = \Delta (f_{2 \bar 2}^\dagger f_{2 \bar 2} + f_{\bar 3 3}^\dagger f_{\bar 3 3})$.  
$\rho_\textrm{I,cuts}$ can have entanglement between B and AC since each excitation at the cuts is mapped onto one of the four Bell type states: $|\text{Bell}\rangle^q$, $|0_{\bar{2}\bar{3}}\rangle^q|1_{23}\rangle^q - i |1_{\bar{2}\bar{3}}\rangle^q |0_{23} \rangle^q$ (excited by $f_{2\bar{2}}^\dagger$), $|0_{\bar{2}\bar{3}}\rangle^q |1_{23}\rangle^q + i |1_{\bar{2}\bar{3}}\rangle^q |0_{23} \rangle^q$ (by $f_{\bar{3}3}^\dagger$), and $|0_{\bar{2}\bar{3}}\rangle^q |0_{23}\rangle^q - i |1_{\bar{2}\bar{3}}\rangle^q|1_{23} \rangle^q$ (by $f_{2\bar{2}}^\dagger f_{\bar{3}3}^\dagger$). 
These Bell states originate from entanglement generation by changing non-Abelian fusion pairs from Fig. ~\ref{system_H1}(b) to (d).  
The second factor $\rho'_\textrm{I}$ accounts for excitations localized in one of B and AC, and does not contribute to the entanglement.
Hence, $\mathcal{LN}(\rho_\textrm{I}) = \mathcal{LN}(\rho_\textrm{I,cuts})$ and $\mathcal{F}(\rho_\textrm{I}) = \mathcal{F}(\rho_\textrm{I,cuts})$. 


$\mathcal{LN}$ and $\mathcal{F}$ are computable since $\rho_\textrm{I, cuts}$ is mapped onto a two-qubit state.  Both show qualitatively the same $T$ dependence in Fig.~\ref{H1_finiteT}.
At $T = 0$, $\mathcal{LN}=\mathcal{F}=1$  because of $|\textrm{Bell} \rangle^q$.
This value is related to the quantum dimension $\sqrt{2}$ of Majoranas; among four ($= \sqrt{2}^4$) electron-number states formed by the four cut Majoranas $\gamma_{2,\bar{2},\bar{3},3}$, only two even-parity states $|0_{\bar{2}\bar{3}} 0_{23} \rangle$ and $|1_{\bar{2}\bar{3}} 1_{23} \rangle$ are allowed at $T=0$ due to superconductivity, resulting in $\ln_2 2 = 1$.

As $T$ increases, the four Bell states become more mixed, so $\mathcal{LN}$ and $\mathcal{F}$ decrease, vanishing at $T_\textrm{SD}$, 
\begin{align}
  k_B T_\textrm{SD} = - \frac{\Delta}{\log (\sqrt 2-1)} \approx 1.135 \Delta. \label{SDtemperature}
\end{align}
Interestingly, $d\mathcal{LN}/dT$ and $d\mathcal{F}/dT$ are discontinuous at $T_\textrm{SD}$, contrary to exponential or algebraic decay; $\mathcal{LN} \propto \delta T$ and $\mathcal{F} \propto (\delta T)^2\log \delta T$ at $T = T_\textrm{SD} - \delta T$ with $\delta T \ll \Delta$, and $\mathcal{LN}, \mathcal{F} = 0$ for $T>T_\textrm{SD}$~\cite{Supple}.
This behavior may be called ``entanglement sudden death''~\cite{Yu09}. It results from thermal mixing of states of different parity in $\rho_\textrm{I}(T)$; it does not occur for a mixture of states of even parity~\cite{Supple}. The overall $T$ dependence is independent of cut positions.

 


By contrast, for the spin chain, $\mathcal{LN} = \mathcal{F} = 0$ at any $T$ as at $T=0$. This indicates that the nonvanishing entanglement in the Kitaev chain is a consequence of the fermion statistics absent in spins.
 
\begin{figure}[bt]
\includegraphics[width=\columnwidth]{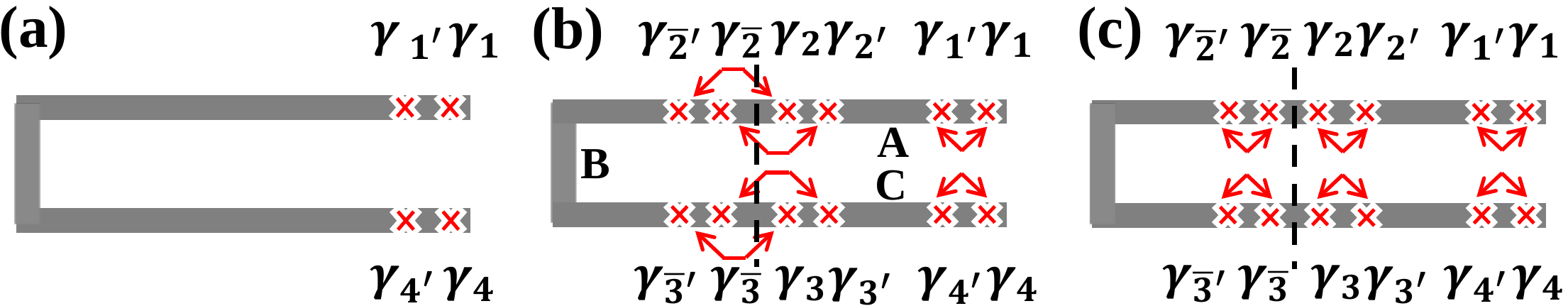}
\caption{ (a) A chain having two Majorana zero modes $\gamma_{a=1,1',4',4}$ at each end. (b) 
It is partitioned as in Fig.~\ref{system_H1}. Four Majoranas $\gamma_{a=2',2, \bar{2}, \bar{2}', \bar{3}', \bar{3} ,3, 3'}$ are revealed at each cut. The arrows indicate the fusion pairs annihilating $|0 \rangle_\textrm{II}$. (c) The Majoranas are fused into fermions localized in B or AC. } \label{system_H2}
\end{figure}

{\it Chain with two Majoranas at an end.---} We next study a chain having two Majorana zero modes $\gamma_{a=1,1',4',4}$ at each end [Fig.~\ref{system_H2}(a)]. It represents a topological phase (D class) of 1D fermions distinct from the Kitaev chain $\hat{H}_\textrm{I}$~\cite{Fidkowski11, Turner11, Wen12}.  Its Hamiltonian is $\hat{H}_\textrm{II} = -\frac{\Delta}{2}\sum_{j=1}^{N-2} (c_j + c_j^\dagger)(c_{j+2} -c_{j+2}^\dagger)$. 
Because of $\gamma_{a=1,1',4',4}$, it has four degenerate ground states $|n \rangle_\textrm{II}$ = $(f_{11'}^\dagger)^{n_{11'}} (f_{44'}^\dagger)^{n_{44'}}|0_{11'} 0_{2 \bar{2}'} 0_{2' \bar{2}} 0_{\bar{3}' 3} 0_{\bar{3} 3'} 0_{44'}\cdots \rangle$ where $n = n_{11'} + 2n_{44'}$ and $n_{11'},n_{44'} = 0,1$.
Changing Majorana fusion pairs from Fig. ~\ref{system_H2}(b) to (c),
we rewrite $|n\rangle_\textrm{II}$ using operators local with respect to regions A, B, C~\cite{Supple}
\begin{align}
|n \rangle_\textrm{II} = & \frac{1}{2} (f_{11'}^\dagger)^{n_{11'}}  (f_{44'}^\dagger)^{n_{44'}} (1 + f_{22'}^\dagger f_{\bar{2}\bar{2}'}^\dagger)(1 + f_{\bar{3}\bar{3}'}^\dagger f_{3 3'}^\dagger) \nn\\
&\quad |0_{11'} 0_{22'} 0_{\bar{2}\bar{2}'} 0_{\bar{3}\bar{3}'} 0_{33'} 0_{44'} \cdots \rangle.
\label{ABCstates2}
\end{align}
Mapping it onto qubits, 
we first observe entanglement between A and BC (AB and C), and find
that the result has a factor of a Bell state $|\textrm{Bell}_2\rangle^q$ ( $|\textrm{Bell}_3\rangle^q$) localized at the cut between A and BC (AB and C)
\begin{align}
|\textrm{Bell}_{a=2,3} \rangle^q &= \frac{1}{\sqrt 2}(|0_{\bar{a}\bar{a}'} \rangle^q |0_{aa'} \rangle^q + |1_{\bar{a}\bar{a}'} \rangle^q |1_{aa'} \rangle^q)   \label{Bell2} 
\end{align}
that entangles qubits $n_{22'}$ and $n_{\bar{2}\bar{2}'}$ ($n_{\bar{3}\bar{3}'}$ and $n_{33'}$).

Entanglement between B and AC is, however, nonlocal due to the fermion statistics. To see this, we need to map 
the ground states into qubits after the subsystem operator grouping such that operators $f_{\bar 2 \bar 2'}^\dagger$, $f_{\bar 3 \bar 3'}^\dagger$ of B are collected to the left of those of AC: $|n \rangle_\textrm{II} = \frac{1}{2}  (1 -  f_{\bar{2}\bar{2}'}^\dagger f_{22'}^\dagger  + f_{\bar{3}\bar{3}'}^\dagger f_{3 3'}^\dagger +f_{\bar{2}\bar{2}'}^\dagger    f_{\bar{3}\bar{3}'}^\dagger f_{22'}^\dagger  f_{3 3'}^\dagger)(f_{11'}^\dagger)^{n_{11'}}  (f_{44'}^\dagger)^{n_{44'}} | 0_{22'} 0_{\bar{2}\bar{2}'} 0_{\bar{3}\bar{3}'} 0_{33'}  \cdots \rangle$,
\begin{align}
|n\rangle_\textrm{II} & \to |n\rangle_\textrm{II}^q = |\textrm{CL}\rangle^q |n_{11'}\rangle^q |n_{44'}\rangle^q,\label{CLentanglement}\\
|\textrm{CL}\rangle^q & = \frac{1}{2} [|0_{\bar{2} \bar{2}'} \rangle^q  |0_{22'} \rangle^q ( |0_{\bar{3}\bar{3}'} \rangle^q |0_{33'} \rangle^q +|1_{\bar{3}\bar{3}'}\rangle^q |1_{33'}  \rangle^q) \nn\\
& \quad- |1_{\bar{2} \bar{2}'}\rangle^q |1_{22'}\rangle^q ( |0_{\bar{3}\bar{3}'}\rangle^q  |0_{33'} \rangle^q -|1_{\bar{3}\bar{3}'}\rangle^q |1_{33'} \rangle^q)]. \nonumber
\end{align}
$|n\rangle_\textrm{II}$ has a factor of a four-qubit cluster state~\cite{Briegel,Guhne} $|\textrm{CL}\rangle^q$. 
$|\textrm{CL}\rangle^q$ is nonlocal as it cannot be written as a product of locally entangled states $|\textrm{Bell}_2 \rangle^q \otimes |\textrm{Bell}_3 \rangle^q$. It is independent of cut positions.
Because the end qubits $|n_{11'} \rangle^q \otimes |n_{44'} \rangle^q$ are decoupled [Eq.~\eqref{CLentanglement}], the equal mixture of ground states $\rho_\textrm{II} (T=0)$  also has the cluster-state entanglement. Since pure excited states have similar cluster-state entanglement, the thermal state $\rho_\textrm{II} (T) = e^{- \beta \hat{H}_\textrm{II}}/\textrm{Tr} e^{- \beta \hat{H}_\textrm{II}}$ 
can have the nonlocal and length-independent entanglement between B and AC.
 

By contrast, for the spin chain obtained by JWT~\cite{Supple} of $\hat{H}_\textrm{II}$, entanglement between B and AC is local.
The equal mixture of ground states $\rho_\textrm{II}^s(T=0)$ has entanglement $|\textrm{Bell}_2 \rangle^s \otimes |\textrm{Bell}_3 \rangle^s$ between B and AC, where $|\textrm{Bell}_2 \rangle^s$ ($|\textrm{Bell}_3 \rangle^s$) is the Bell state in Eq.~\eqref{Bell2} localized at the cut between A and BC (AB and C). 
Similarly, the entanglement is local in all the excited states. 
Consequently, the thermal state $\rho_\textrm{II}^s(T)$ of the spin chain is decomposed as $\rho_\textrm{II}^s(T) = \rho_\textrm{II,cut2}^s(T)\otimes \rho_\textrm{II,cut3}^s(T) \otimes {\rho_\textrm{II}^s}'(T)$,
where $\rho_\textrm{II,cut2}^s$ ($\rho_\textrm{II,cut3}^s$) can have local entanglement at the cut between A and BC (AB and C) and ${\rho_\textrm{II}^s}'$ has no entanglement.
Since $\rho^s_\textrm{II}(T)$ is identical to the result of mapping  $\rho_\textrm{II}(T)$ into qubits without the subsystem operator grouping, we see that the nonlocal entanglement in the electron state $\rho_\textrm{II}(T)$ derives from the fermion exchange statistics. 



\begin{figure}[bt]
\includegraphics[width=\columnwidth]{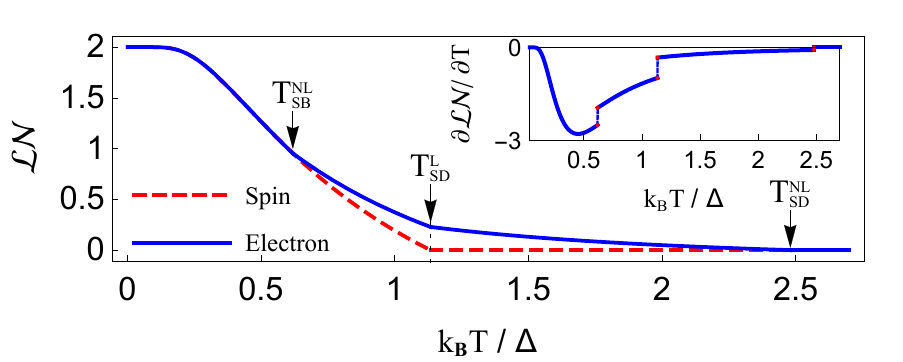}
\caption{Temperature dependence of entanglement ($\mathcal{LN}$) between B and AC for the cases of the electron chain of $\hat{H}_\textrm{II}$ and the corresponding spin chain.
The electrons have $\mathcal{LN}$ greater than or equal to the spins. The excess indicates nonlocality of the entanglement in the electrons as the entanglement is local in the spins. The nonlocality is visible in $T \in (T_\textrm{SB}^\textrm{NL}, T^\textrm{NL}_\textrm{SD})$.
Inset: $d \mathcal{LN} / dT$ is discontinuous at $T_\textrm{SB}^\textrm{NL}$, $T_\textrm{SD}^\textrm{L}$, $T^\textrm{NL}_\textrm{SD}$.
}
\label{H2_finiteT}
\end{figure}

We compute $\mathcal{LN}$ for the electron state $\rho_\textrm{II}(T)$ and the spin $\rho^s_\textrm{II}(T)$ in Fig.~\ref{H2_finiteT}; $\mathcal{F}$ is hard to compute for four-qubit mixed states~\cite{Lee12}. For electrons, 
$\mc {LN}(\rho_\textrm{II})= \mc{LN}(\rho_\textrm{II,cuts})$ where $\rho_\textrm{II,cuts} \propto e^{-\beta \hat H_{\textrm{II,cuts}}}$ and $\hat H_\textrm{II,cuts} = \Delta (f_{22'} f_{\bar 2 \bar 2'}+f_{\bar 3 \bar 3'} f_{33'}+\textrm{H.c})$. In contrast, for spins, the decomposition of $\rho^s_\textrm{II}$ above implies $\mathcal{LN}(\rho^s_\textrm{II}) = \mathcal{LN}(\rho^s_\textrm{II,cut2})+ \mathcal{LN}(\rho^s_\textrm{II,cut3}) = 2 \mathcal{LN}(\rho^s_\textrm{II,cut2})$.

For both $\rho_\textrm{II}(T=0)$ and $\rho^s_\textrm{II}(T=0)$, $\mathcal{LN}$ equals $2$ and decreases at larger $T$. $\mathcal{LN}$ does not distinguish between the nonlocal entanglement of $\rho_\textrm{II}(T=0)$ and the local one of $\rho^s_\textrm{II}(T=0)$.
  However, it distinguishes at finite $T$ ($ > T_{\text{SB}}^{\text{NL}}$), showing a larger value for $\rho_\textrm{II}(T)$.
The excess $\mathcal{LN}(\rho_\textrm{II}(T)) - \mathcal{LN}(\rho^s_\textrm{II}(T))$ is a marker of
the breakdown of the fermion-spin correspondence, i.e.,
the nonlocality of entanglement between B and AC in $\rho_\textrm{II}(T)$.



Interestingly, the excess starts to appear at $T = T_\textrm{SB}^\textrm{NL}$, increases with $T$ till $T^\textrm{L}_\textrm{SD}$ (at which the local entanglement of $\rho^s_\textrm{II}$ vanishes), then decreases with $T$, vanishing at $T^\textrm{NL}_\textrm{SD}$. $d \mathcal{LN} / dT$ is discontinuous at $T_\textrm{SB}^\textrm{NL}$, $T^\textrm{L}_\textrm{SD}$, $T^\textrm{NL}_\textrm{SD}$, which are related to entanglement sudden birth and death~\cite{Yu09}. 
It is nontrivial that the quantum nonlocality induced by the fermion statistics is more visible at finite $T$. This may be because the marker is based on a bipartite entanglement measure rather than a multipartite (4-qubit) one.

{\it Conclusion.---} 
We find nonlocal entanglement in 1D electrons having end Majoranas. 
It is independent of subsystem lengths, occurs in the bulk with protection by energy gap, 
and survives up to certain temperature. Its form (Bell state vs. cluster state) and its finite-temperature behavior depend on the number of the end Majoranas. This is an entanglement version of bulk-edge correspondence for the 1D electrons.  
 
The nonlocal entanglement will exist universally in general 1D fermions belonging to BDI and D classes, provided that their subsystems A, B, C are longer than Majorana localizaion length $\xi$. ($\xi$ was zero in our models.)
It is because the nonlocal entanglement originates from Majorana non-Abelian fusion and the fermion statistics,  not depending on system specifics. It will be interesting to rigorously confirm this universality.



Our results illustrate a number of interesting aspects of fermion entanglement.
First, the fermion statistics plays a crucial role in entanglement, especially for mixed states: It can generate quantum nonlocality.
Due to this, the fermion-spin correspondence 
established~\cite{Chen,Bahri14} for pure states breaks down: 
The thermal state  $\rho_\textrm{I}$ of electrons ($\rho_\textrm{I}^s$ of spins) has nonlocal (no) entanglement.
For the thermal state $\rho_\textrm{II}$ of electrons ($\rho_\textrm{II}^s$ of spins), the amount of entanglement between B and AC can differ from (always equals) the sum of local entanglement at the two cuts.
 Even for pure states, the form of entanglement can differ between the two connected by JWT; e.g., cluster in $|n \rangle_\textrm{II}$ vs. $\textrm{Bell}\otimes\textrm{Bell}$ in $|n \rangle_\textrm{II}^s$.

 
Second, bipartite mixed-state entanglement measures unveil nontrivial temperature dependence of the nonlocal entanglement. Their sudden birth and death provide a characteristic example of finite-temperature behavior of topological phases, which is in stark contrast with usual quantum-to-classical crossover.
 

Third, the electron thermal states are mixtures of states of different parity. For the case where the parity mixing is not allowed for some purposes, e.g., where only states of even parity are thermally mixed, our results of $\mathcal{LN}$ and $\mathcal{F}$ provide a lower bound of the entanglement~\cite{Banuls07,Supple}. Hence for that case, nonlocal entanglement is also generated by the fermion statistics.
 

Fourth, our finding will be useful for generalizing and characterizing the notion of topological phases to thermal states, overcoming difficulties faced by tools for pure states such as ground-state degeneracy, bulk-edge correspondence, topological entanglement entropy~\cite{Kitaev06,Levin06}.





We thank E. Berg, R. Fazio, P. Fendley, F. Pollmann, and S. Ryu for valuable discussions, and the support by Korea NRF (Grant Nos. 2015R1A2A1A15051869 and 2016R1A5A1008184).


\begin{thebibliography}{99}

\bibitem{kitaev01} A. Yu. Kitaev, 
{Unpaired Majorana fermions in quantum wires},
{\it Physics-Uspekhi} {\bf 44}, 131 (2001).

\bibitem{Fidkowski11} L. Fidkowski,  and A. Yu. Kitaev, 
{Topological phases of fermions in one dimension},
{\it Phys. Rev. B} {\bf 83}, 075103 (2011).

\bibitem{Turner11} A. M. Turner, F. Pollmann, and E. Berg, 
{Topological phases of one-dimensional fermions: An entanglement point of view},
{\it Phys. Rev. B} {\bf 83}, 075102 (2011).

\bibitem{Wen12} X.-G. Wen,
{Symmetry-protected topological phases in noninteracting fermion systems,}
{\it Phys. Rev. B} {\bf 85}, 085103 (2012).

\bibitem{Kim14} Isaac H. Kim,
{Entropic topological invariant for a gapped one-dimensional system},
{\it Phys. Rev. B} {\bf 89}, 235120 (2014).

\bibitem{Bahri14} Y. Bahri and A. Vishwanath,
Detecting Majorana fermions in quasi-one-dimensional topological phases using nonlocal order parameters,
{\it Phys. Rev. B} {\bf 89}, 155135 (2014).

\bibitem{Kitaev03} A. Yu. Kitaev, 
{Fault-tolerant quantum computation by anyons,}
{\it Ann. Phys.} {\bf 303}(1), 2-30 (2003).

\bibitem{Alicea11} J. Alicea, Y. Oreg, G. Refael, F. von Oppen, and M. P. A. Fisher, 
{Non-Abelian statistics and topological quantum information processing in 1D wire networks,}
{\it Nat. Phys.} {\bf 7}, 412 (2011).



\bibitem{Castelnovo07} C. Castelnovo and C. Chamon,
{Entanglement and topological entropy of the toric code at finite temperature,}
{\it Phys. Rev. B} {\bf 76}, 184442 (2007).

\bibitem{Nussinov08} Z. Nussinov and G. Ortiz,
Autocorrelations and thermal fragility of anyonic loops in topologically quantum ordered systems,
{\it Phys. Rev. B} {\bf 77}, 064302 (2008).

\bibitem{Hastings11} M. B. Hastings, 
{ Topological order at nonzero temperature,}
{\it Phys. Rev. Lett.} {\bf 107}, 210501 (2011).

\bibitem{Poulin13} O. Landon-Cardinal and D. Poulin, 
{Local topological order inhibits thermal stability in 2D,}
{\it Phys. Rev. Lett.} {\bf 110}, 090502 (2013).

\bibitem{Viyuela14} O. Viyuela, A. Rivas, and M. A. Martin-Delgado, 
{Uhlmann Phase as a Topological Measure for One-Dimensional Fermion Systems,}
{\it Phys. Rev. Lett.} {\bf 112}, 130401 (2014); O. Viyuela, A. Rivas, and M. A. Martin-Delgado, {Two-dimensional density-matrix topological fermionic phases: topological Uhlmann numbers,} {\it Phys. Rev. Lett.}, {\bf 113}, 076408 (2014).

\bibitem{Arovas14} Z. Huang and D. P. Arovas,
{Topological Indices for Open and Thermal Systems via Uhlmann's Phase,}
{\it Phys. Rev. Lett.} {\bf 113}, 076407 (2014).

\bibitem{Huber14} E. P. L. van Nieuwenburg, and S. D. Huber, 
{Classification of mixed-state topology in one dimension,}
{\it Phys. Rev. B} {\bf 90}, 075141 (2014).

\bibitem{Budich15} J. C. Budich and S. Diehl,
{Topology of density matrices,}
{\it Phys. Rev. B} {\bf 91}, 165140 (2015).

\bibitem{Grusdt16} F. Grusdt,
{Topological order of mixed states in quantum many-body systems,}
arXiv:1609.02432 (2016).



\bibitem{Pollmann10} F. Pollmann, A. M. Turner, E. Berg, and M. Oshikawa,
{Entanglement spectrum of a topological phase in one dimension,}
{\it Phys. Rev. B} {\bf 81}, 064439 (2010).

\bibitem{Li08} Hui Li and F. D. M. Haldane,
{Entanglement spectrum as a generalization of entanglement entropy: Identification of topological order in non-abelian fractional quantum hall effect states,}
{\it Phys. Rev. Lett.} {\bf 101}, 010504 (2008).

\bibitem{Fidkowski10} L. Fidkowski,
{Entanglement spectrum of topological insulators and superconductors,}
{\it Phys. Rev. Lett.} {\bf 104}, 130502 (2010).

\bibitem{Chen} X. Chen, Z.-C. Gu, and X.-G. Wen,
Complete classification of one-dimensional gapped quantum phases in interacting spin systems,
{\it Phys. Rev. B} {\bf 84}, 235128 (2011).

\bibitem{Plenio07} M. B. Plenio and S. Virmani, {An introduction to entanglement measures},
  {\it Quant. Inf. Comp.} {\bf 7}, 1 (2007).
  
\bibitem{Bennett96}
C. H. Bennett, D. P. DiVincenzo, J. A. Smolin, and W. K. Wootters, 
{Mixed-state entanglement and quantum error correction},
{\it Phys. Rev. A} {\bf 54}, 3824 (1996).
  
\bibitem{Lee00} J. Lee, M. S. Kim, Y. J. Park, and S. Lee,
 {Partial teleportation of entanglement in a noisy environment,}
  {\it J. Mod. Opt.} {\bf 47} (12), 2151-2164 (2000).

\bibitem{Vidal02} G. Vidal and R. F. Werner, 
{Computable measure of entanglement},
{\it Phys. Rev. A} {\bf 65}, 032314 (2002).

\bibitem{Plenio05} M. B. Plenio, 
{Logarithmic negativity: a full entanglement monotone that is not convex,} 
{\it Phys. Rev. Lett.} {\bf 95}, 090503 (2005).

\bibitem{Other} The cases of $\mu \ne 0$ and $t \ne \Delta$ will be reported elsewhere.

\bibitem{Supple} See Supplementary Material, which includes Refs.~\cite{Vollbrecht01, Wootters98}.

\bibitem{Zanardi02} P. Zanardi,
Quantum entanglement in fermionic lattices,
{\it Phys. Rev. A} {\bf 65}, 042101 (2002).

\bibitem{Banuls07} M.-C. Ba\~{n}uls, J. I. Cirac, and M. M. Wolf,
{Entanglement in fermionic systems}, 
{\it Phys. Rev. A} {\bf 76}, 022311 (2007).

\bibitem{Ivanov01} D. A. Ivanov,
Non-Abelian statistics of half-quantum vortices in p-wave superconductors,
{\it Phys. Rev. Lett.} {\bf 86}, 268 (2001).

\bibitem{Goldstein11} G. Goldstein and C. Chamon,
Decay rates for topological memories encoded with Majorana fermions,
{\it Phys. Rev. B} {\bf 84}, 205109 (2011).

\bibitem{Budich12} J. C. Budich, S. Walter, and B. Trauzettel,
Failure of protection of Majorana based qubits against decoherence,
{\it Phys. Rev. B} {\bf 85}, 121405(R) (2012).

\bibitem{Rainis12} D. Rainis and D. Loss,
Majorana qubit decoherence by quasiparticle poisoning,
{\it Phys. Rev. B} {\bf 85}, 174533 (2012).

\bibitem{Montero11} M. Montero and E. Mart\'{i}n-Mart\'{i}nez,
Fermionic entanglement ambiguity in noninertial frames,
{\it Phys. Rev. A} {\bf 83}, 062323 (2011).

\bibitem{Friis13} N. Friis, A. R. Lee, and D. E. Bruschi,
Fermionic-mode entanglement in quantum information,
{\it Phys. Rev. A} {\bf 87}, 022338 (2013). 

\bibitem{Bayat12} A. Bayat, S. Bose, P. Sodano, and H. Johannesson, Entanglement Probe of Two-Impurity Kondo Physics in a Spin Chain, {\it Phys.
Rev. Lett.} {\bf 109}, 066403 (2012). 

\bibitem{YALee13} Y. A. Lee and G. Vidal,
{Entanglement negativity and topological order,}
{\it Phys. Rev. A} {\bf 88}, 042318 (2013).

\bibitem{Castelnovo13} C. Castelnovo,
{Negativity and topological order in the toric code,}
{\it Phys. Rev. A} {\bf 88}, 042319 (2013).

\bibitem{Calabrese15} P. Calabrese, J. Cardy, and E. Tonni,
{Finite temperature entanglement negativity in conformal field theory,}
{\it J. of Phys. A: Mathematical and Theoretical} {\bf 48}, 015006 (2015).

\bibitem{Lee15} S.-S. B. Lee, J. Park, and H.-S. Sim, 
{Macroscopic Quantum Entanglement of a Kondo Cloud at Finite Temperature,}
{\it Phys. Rev. Lett.} {\bf 114}, 057203 (2015).

\bibitem{Eisler} V. Eisler and Z. Zimboras, On the partial transpose of fermionic Gaussian states,
{\it New J. of Phys.} {\bf 17}, 053048 (2015); J. Eisert, V. Eisler, and Z. Zimboras, Entanglement negativity bounds for fermionic Gaussian states, Arxiv:1611.08007 (2016).

\bibitem{Shapourian} H. Shapourian, K. Shiozaki, and S. Ryu, Partial time-reversal transformation and entanglement negativity in fermionic systems, Arxiv:1611.07536 (2016).

\bibitem{Yu09} Yu, T. \& Eberly, J. H.
{Sudden death of entanglement.}
{\it Science} {\bf 323}, 598 (2009).

\bibitem{Briegel} H. J. Briegel and R. Raussendorf, Persistent entanglement in arrays of interacting particles, {\it Phys. Rev. Lett.} {\bf 86}, 910 (2001).

\bibitem{Guhne} O. G\"{u}hne and G. T\'{o}th, Entanglement detection, {\it Phys. Rep.} {\bf 474}, 1 (2009).

\bibitem{Lee12} S.-S. B. Lee and H.-S. Sim, Quantifying mixed-state quantum entanglement by optimal entanglement witnesses, {\it Phys. Rev. A} {\bf 85}, 022325 (2012).

\bibitem{Kitaev06} A. Kitaev and J. Preskill, 
{Topological Entanglement Entropy,}
{\it Phys. Rev. Lett.} {\bf 96}, 110404 (2006).

\bibitem{Levin06} M. Levin, and X.-G. Wen, 
{Detecting Topological Order in a Ground State Wave Function,}
{\it Phys. Rev. Lett.} {\bf 96}, 110405 (2006).

\bibitem{Vollbrecht01} K. G. H. Vollbrecht and R. F. Werner,
Entanglement measures under symmetry.
{\it Phys. Rev. A} {\bf 64}, 062307 (2001).

\bibitem{Wootters98} W. K. Wootters,
Entanglement of formation of an arbitrary state of two qubits.
{\it Phys. Rev. Lett.} {\bf 80}, 2245 (1998).

\end{thebibliography}
\end{document}